\documentclass[
aps, prx, 
preprint,
lengthcheck,
superscriptaddress,
nofootinbib,
nobibnotes,
]{main-revtex4-2}

\usepackage{amsmath, amsfonts, amssymb, bbm}
\usepackage{bm}
\usepackage{mathtools}
\usepackage{arydshln}   
\usepackage{slashed}
\usepackage{enumerate} 
\usepackage{graphicx} 
\usepackage{subfigure} 
\usepackage[percent]{overpic}
\usepackage{color}
\usepackage{multirow}
\usepackage{booktabs}

\definecolor{mainPurple}{RGB}{126, 30, 156}
\definecolor{mainViolet}{RGB}{154, 14, 234}
\definecolor{mainOrange}{RGB}{249, 115, 6}
\definecolor{mainSlateblue}{RGB}{0, 0, 139}

\usepackage{hyperref}

\newcommand{\addrUCAS}{University of Chinese Academy of Sciences (UCAS), Beijing 100049, China
}
\newcommand{\addrIctpAp}{International Centre for Theoretical Physics Asia-Pacific (ICTP-AP), 
University of Chinese Academy of Sciences (UCAS), Beijing, 100190, China
}
\newcommand{\addrTaijiLabBeijing}{
Taiji Laboratory for Gravitational Wave Universe (Beijing/Hangzhou), 
University of Chinese Academy of Sciences (UCAS), Beijing, 100049, China
}

\begin{document}

\title{Test Gravitational-Wave Polarizations with Space-Based Detectors}

\author{Jun-Shuai Wang}
\email{wangjunshuai@ucas.ac.cn}
\affiliation{\addrIctpAp}
\affiliation{\addrTaijiLabBeijing}
\affiliation{\addrUCAS}

\author{Chang Liu}
\email{liuchang@yzu.edu.cn}
\affiliation{Center for Gravitation and Cosmology, College of Physical Science and Technology, Yangzhou University, Yangzhou, 225009, China}

\author{Ju Chen}
\email{chenju@ucas.ac.cn}
\affiliation{\addrIctpAp}
\affiliation{\addrTaijiLabBeijing}
\affiliation{\addrUCAS}

\author{Jibo He}
\email{jibo.he@ucas.ac.cn}
\affiliation{\addrTaijiLabBeijing}
\affiliation{\addrUCAS}
\affiliation{Hangzhou Institute for Advanced Study, UCAS, Hangzhou 310024, China}
\date{\today}

\begin{abstract}
In this work, we systematically investigate the capability of space-based gravitational wave detectors in constraining parameters of non-tensor polarization modes. 
Using Bayesian inference and Fisher Information Matrix methods, we analyze gravitational wave signals from the inspiral phase of supermassive binary black hole mergers. 
By starting with time-domain signals and applying Fourier transforms, we avoid the use of the stationary phase approximation. 
We found an asymmetry in the estimation of the vector-mode parameter $\alpha_x$ at inclination angles $\iota = 0$ and $\iota = \pi$, which has not been explicitly pointed out in previous studies. 
We also observe strong correlations between scalar-mode parameters, $\alpha_b$ and $\alpha_l$, which currently limit their independent estimation. 
These findings underscore the importance of using complete inspiral-merger-ringdown waveforms to enhance the ability to distinguish the non-tensor polarization modes. 
Finally, we employ a new LISA-Taiji network configuration, in which the orientation of spacecrafts of Taiji maintains a fixed phase offset relative to these of LISA. 
Under the adiabatic approximation and the assumption of equal arms, this phase is
found to have no significant effect on data analysis.

\end{abstract}

\maketitle

\section{Introduction}
\label{sec:intro}

Gravitational waves (GWs) are ripples in spacetime predicted by Einstein's general theory of relativity, arising from the violent motion of massive objects, such as the merger of binary black holes or binary neutron stars. 
Since the first direct detection of GWs by LIGO in 2015~\cite{abbottObservationGravitationalWaves2016a}, gravitational wave astronomy has transitioned from theoretical verification to practical observation, 
which enabled us to test general relativity in the strong-field regime~\cite{LIGOScientific:2016aoc, LIGOScientific:2019fpa, Stairs:2003eg, Will:2014kxa}.

While general relativity predicts that GWs propagates only through tensor polarizations (the plus and cross modes), alternative theories of gravity suggest up to six polarization modes~\cite{Eardley:1973zuo, Eardley:1973br}. 
For example, Brans-Dicke theory predicts an additional scalar polarization mode beyond the tensor modes~\cite{Brans:1961sx}; f(R) gravity includes two addition scalar polarization modes~\cite{Alves:2009eg, RizwanaKausar:2016zgi, Liang:2017ahj, Niu:2019ywx}; Einstein-Aether theory suggests the existence of both scalar and vector polarization modes~\cite{Jacobson:2004ts, Lin:2018ken, Zhang:2019iim}, and generalized tensor-vector-scalar theories, like TeVeS theory, predict that all six polarization modes could exist~\cite{Bekenstein:2004ne, Gong:2018ybk}. 
Therefore, detecting these non-tensor modes provides a crucial way to test general relativity and examine various alternative theories of gravity that beyond general relativity~\cite{LIGOScientific:2018czr}.

Ground-based detectors, such as LIGO, Virgo, and KAGRA, have achieved remarkable success in observing tensor modes. 
A network of ground-based detectors, consisting of LIGO, Virgo, KAGRA, and LIGO India, will have the capability to detect additional polarization modes~\cite{Yunes:2013dva, Hagihara:2019ihn, Takeda:2019gwk}. 
However, their limited baseline lengths and arrangement configurations restrict sensitivity to subdominant non-tensor components~\cite{isi2017probinggravitationalwavepolarizations, Takeda:2018uai}. 
There are several challenges in the current detection of non-tensor modes: first, the duration of the signal  in ground-based detection is extremely short, often lasting only a few seconds; second, the predicted non-tensor waveforms vary significantly between different modified gravity theories, current research depends on parameterized framework (ppE framework)~\cite{Yunes:2009ke, Chatziioannou:2012rf, Chatziioannou:2012rf1}. 
Although it cannot parameterize all possible types of deviations from general relativity, its scope of applicability has been discussed in literature~\cite{Yunes:2013dva}. 
In current data analysis, only the inspiral phase signals can be used for testing non-tensor modes~\cite{Hilborn:2018rio}, as the waveforms for the merger and ringdown phases dependent on numerical GR to predict the polarization modes for those phases.

In contrast, space-based detectors like LISA~\cite{amaroseoane2017laserinterferometerspaceantenna} and Taiji~\cite{Hu:2017mde}, which consist of a triangular configuration formed by three spacecrafts orbiting the Sun, have interferometric arm lengths at the order of millions of kilometers and dynamic orbital configurations. 
These features provide them with highly promising for detecting additional polarization modes of gravitational waves~\cite{Gair:2012nm, Philippoz:2017ywb}, across a wider frequency bandwidth ranging from 0.1~mHz to 1~Hz, which enables long-duration detection of non-tensor radiation from persistent sources (such as Galactic double white dwarfs) or the mergers of supermassive black holes. 
The detector's response to each polarization mode varies with the orbital position, which can enhance the ability to separate different polarization modes. 
Space-based detector networks have more degrees of freedom in formation configuration compared to ground-based networks. 
Through optimized arrangements, the sensitivity to non-tensor modes can be significantly improved~\cite{wangAlternativeLISATAIJINetworks2021}.

Although additional polarization modes can be tested through transient burst GWs~\cite{hayamaModelindependentTestGravity2013}, continuous gravitational waves~\cite{isiDetectingBeyondEinsteinPolarizations2015}, and stochastic gravitational wave backgrounds~\cite{Nishizawa:2009bf, Callister:2017ocg}, in this paper we focus on investigating non-tensor polarization modes in the context of the inspiral phase of supermassive binary black holes. 


In the previous works which assuming that non-tensor modes are dominated by dipole radiation, the frequency-domain waveform is obtained by using the stationary phase approximation. 
While we start from the time-domain waveform and obtain the frequency-domain waveform through Fourier transform, this approach naturally satisfies the requirement that the frequency of non-tensor modes is half that of tensor modes.
In the existing results, the variation of the vector mode coefficient $\alpha_x$ with $\iota$ appears to be symmetric. 
However, we actually find that it is not strictly symmetric. This aspect will be discussed in detail in Sect.~\ref{sec:sub:iota}.
The Fisher information matrix is widely used due to its computational efficiency. In our work, we provide the results of Bayesian estimation and Fisher estimation, simultaneously.

We have found that there have been numerous studies on the networking of space detectors, all of which are based on the configuration proposed in 2021~\cite{wangAlternativeLISATAIJINetworks2021, wangAlternativeLISATAIJINetworks2021a}. 
In this paper, we investigate the sensitivity of non-tensor mode parameters under a new configuration.

This paper is organized as follows:
In Sect.~\ref{sec:intro}, we present the introduction and research motivation.
In Sect.~\ref{sec:model}, we give a brief overview of the waveform and signal model.
In Sect.~\ref{sec:results}, we present the main results of this work.
In Sect.~\ref{sec:summary}, we provide a summary and discuss possible future improvements.

\section{Waveform and signal model}
\label{sec:model}

Generally, a four-dimensional metric gravity theory has six degrees of freedom~\cite{Eardley:1973br}: 
these consist of two tensor modes (plus and cross), two vector modes ($x$ and $y$), and two scalar modes (breathing and longitudinal).
The perturbed metric ${h}_{ij}(t)$ could be decomposed into the six polarization modes as:
\begin{align}
    \begin{aligned}
        {h}_{ij}(t) = & h_{+}(t) {e}_{ij}^{+} + h_{\times}(t) {e}_{ij}^{\times}
        + h_{x}(t) {e}_{ij}^{x} \\
         &+h_{y}(t) {e}_{ij}^{y}
        + h_{b}(t) {e}_{ij}^{b} + h_{l}(t) {e}_{ij}^{l}
    \end{aligned}
    \label{eq:h_ij}
\end{align}
where $A \in (+, \times, x, y, b, l)$, are symbols representing different polarization modes, 
$h_A$ are waveforms of the six polarization modes, and ${e}_{ij}^{A}$ are the polarization tensors.

\subsection{Waveform}

In this paper, we adopt the heliocentric (sun-mass-center-based) right-handed coordinate system~\cite{Rubbo:2003ap}.
We assume that quadrupole radiation dominates in the tensor modes $(+, \times)$, and the non-tensor polarizations $(x, y, b, l)$ are dominated by dipole radiation, and the leading order of the waveforms can be expressed within the parameterized post-Einsteinian (ppE) framework as~\cite{Chatziioannou:2012rf, OBeirne:2019lwp}:
\begin{align}
    \begin{aligned}
        h_+ &= \frac{4\mathcal{M}}{D_L} (\mathcal{M}\omega)^{2/3} \frac{(1 + \cos^2 \iota)}{2} \cos (2\Phi + 2\Phi_0), \\
        h_\times &= \frac{4\mathcal{M}}{D_L} (\mathcal{M}\omega)^{2/3} \cos \iota \sin (2\Phi + 2\Phi_0), \\
        h_x &= \frac{\alpha_x \mathcal{M}}{D_L} (\mathcal{M}\omega)^{1/3} \cos \iota \cos (\Phi + \Phi_0), \\
        h_y &= \frac{\alpha_y \mathcal{M}}{D_L} (\mathcal{M}\omega)^{1/3} \sin (\Phi + \Phi_0), \\
        h_b &= \frac{\alpha_b \mathcal{M}}{D_L} (\mathcal{M}\omega)^{1/3} \sin \iota \cos (\Phi + \Phi_0), \\
        h_l &= \frac{\alpha_l \mathcal{M}}{D_L} (\mathcal{M}\omega)^{1/3} \sin \iota \cos (\Phi + \Phi_0),
    \end{aligned}
    \label{eq:waveforms_ppE_2}
\end{align}
where $\alpha_x$, $\alpha_y$, $\alpha_b$, $\alpha_l$ are the dimensionless ppE parameters, corresponding to the amplitudes of the vector polarization modes $x$ and $y$, and the scalar breathing mode and longitudinal mode, respectively.
Due to the rotational symmetry between the vector modes $x$ and $y$, one has $\alpha_x=\alpha_y$.
$D_{\rm L}$ is the luminosity distance, $\mathcal{M}=\eta^{3/5} M$ is the chirp mass, $M=m_1 + m_2$ is the total mass and $\eta=m_1 m_2/M^2$ is the symmetry mass ratio.
$\omega=\pi f$ is the orbital angular frequency with the gravitational wave frequency $f$, $\Phi$ is the orbital phase and $\Phi_0$ is the initial phase, $\iota$ is the inclination angle: the angle between the direction of angular momentum of the source and the direction of GW propagation. 

From Eq.~\eqref{eq:waveforms_ppE_2}, we know that the gravitational wave frequencies of the non-tensor modes are half of that of the tensor modes, because of that we assume that the non-tensor polarization modes are dominated by dipole radiation. And this is also the case for the evolution of the orbital phase.

Following Refs.~\cite{OBeirne:2019lwp, Liu:2020mab}, the leading order of the formula describing the evolution of orbital angular frequency $\omega(t)$ is given by:
\begin{align}
    \begin{aligned}
        \frac{d\omega}{dt} =& \alpha_D \omega^3 + \alpha_Q \omega^{11/3}, \\
        \alpha_D =& \alpha_d \eta^{2/5} \mathcal{M},\quad \alpha_Q = \alpha_q \mathcal{M}^{5/3},
    \end{aligned}
\end{align}
where $\alpha_d$ and $\alpha_q$ are the dimensionless ppE parameters that scale the contributions of the dipole and quadrupole components to the frequency evolution, respectively. 
In general relativity, one have $\alpha_x=\alpha_b=\alpha_l = 0$, $\alpha_d=0$ and $\alpha_q=96/5$. 

When $\alpha_d$ is not equal to zero, the relationship between orbital angular frequency $\omega(t)$ and time $t$ can be expressed as:
\begin{align}
    t &= t_0 + \int_{\omega_0}^{\omega_t} \frac{d\omega}{\alpha_D \omega^3 + \alpha_Q \omega^{11/3}} 
    \nonumber\\
    &= t_0 + \frac{1}{4 \alpha_D^4} \Bigg\{ 6 \alpha_Q^2 \alpha_D \bigg( \frac{1}{\omega_0^{2/3}} - \frac{1}{\omega_t^{2/3}} \bigg)
    \nonumber\\
    &\quad + 2 \alpha_D^3 \bigg( \frac{1}{\omega_0^2} - \frac{1}{\omega_t^2} \bigg) + 3 \alpha_Q \alpha_D^2 \bigg( \frac{1}{\omega_t^{4/3}} - \frac{1}{\omega_0^{4/3}} \bigg) \nonumber\\
    & \hspace{-4 mm} + 6 \alpha_Q^3 \bigg[ \log \bigg( 1 + \frac{\alpha_D}{\alpha_Q \omega_t^{2/3}}\bigg) - 
    \log \bigg( 1 + \frac{\alpha_D}{\alpha_Q \omega_0^{2/3}}\bigg) \bigg] \Bigg\},
    \label{eq:omega_time}
\end{align}
Within a given segment of time series $(t_0 \rightarrow t)$, $t_0$ is the initial time, and $t$ is the final time; $\omega_0$ and $\omega_t$ represent the initial and final orbital angular frequencies, respectively. When $\alpha_D$ is sufficiently small, $\alpha_D \ll \alpha_Q \omega_0^{2/3} $, one can obtain the following approximate expression:
\begin{align}
    t =& t_0 + \frac{3}{8}\frac{1}{\alpha_Q}\left( \omega_0^{-\frac{8}{3}}-\omega_t^{-\frac{8}{3}} \right) + \frac{3}{10}\frac{\alpha_D}{\alpha_Q^2} \left( \omega_t^{-\frac{10}{3}}-\omega_0^{-\frac{10}{3}} \right) \nonumber\\
    & + \frac{1}{4}\frac{\alpha_D^2}{\alpha_Q^3} \left( \omega_0^{-4}-\omega_t^{-4} \right) + O(\alpha_D^3).
    \label{eq:omega_time_approx}
\end{align}

The orbital phase $\Phi$ can then be determined by integrating $\omega(t)$ with respect to $t$:
\begin{align}
    \Phi(t) &= \Phi_0 + \int_{t_0}^{t} \omega(t) dt \nonumber\\
    &= \Phi_0 + \int_{\omega_0}^{\omega_t} \frac{d\omega}{\alpha_D \omega^2 + \alpha_Q \omega^{8/3}} \nonumber\\
    &= \Phi_0 + \frac{1}{\alpha_D} \left( \frac{1}{\omega_0} - \frac{1}{\omega_t} \right) + \frac{3 \alpha_Q}{\alpha_D^2} \left( \frac{1}{\omega_t^{1/3}} - \frac{1}{\omega_0^{1/3}} \right) \nonumber\\
    &\quad + 3 \frac{\alpha_Q^{3/2}}{\alpha_D^{5/2}} \left[ \text{arccot} \left( \sqrt{ \frac{\alpha_Q \omega_0^{2/3}}{ \alpha_D }} \right) \right. \nonumber\\
    & \hspace{86 pt} \left. - \text{arccot} \left( \sqrt{\frac{\alpha_Q \omega_t^{2/3}}{\alpha_D}} \right) \right],
\end{align}
where $\Phi_0$ is the orbital phase at time $t_0$, $\Phi(t)$ is the orbital phase at time $t$. 
When $\alpha_D$ is sufficiently small, one can obtain the following approximate expression:
\begin{align}
    \Phi(t) =& \Phi_0 + \frac{3}{5 \alpha_Q} \left( {\omega_0^{-\frac{5}{3}}} - {\omega_t^{-\frac{5}{3}}} \right) - \frac{3 \alpha_D}{7 \alpha_Q^2} \left( {\omega_0^{-\frac{7}{3}}} - {\omega_t^{-\frac{7}{3}}} \right) \nonumber\\
    &+ \frac{\alpha_D^2}{3\alpha_Q^3} \left( {\omega_0^{-3}} - {\omega_t^{-3}} \right) + O(\alpha_D^3).
\end{align}

The paerameters $\alpha_d$ and $\alpha_q$ may be dependent on $\alpha_x$, $\alpha_b$, $\alpha_l$ in some specific theories~\cite{Chatziioannou:2012rf}. 
In this paper, we assume the ppE parameters ($\alpha_d$, $\alpha_q$, $\alpha_x$, $\alpha_b$, $\alpha_l$) are independent for simplicity.

\subsection{Polarization Tensors}

\begin{figure}[tbp]
    \centering
    \includegraphics[width=0.7 \linewidth]{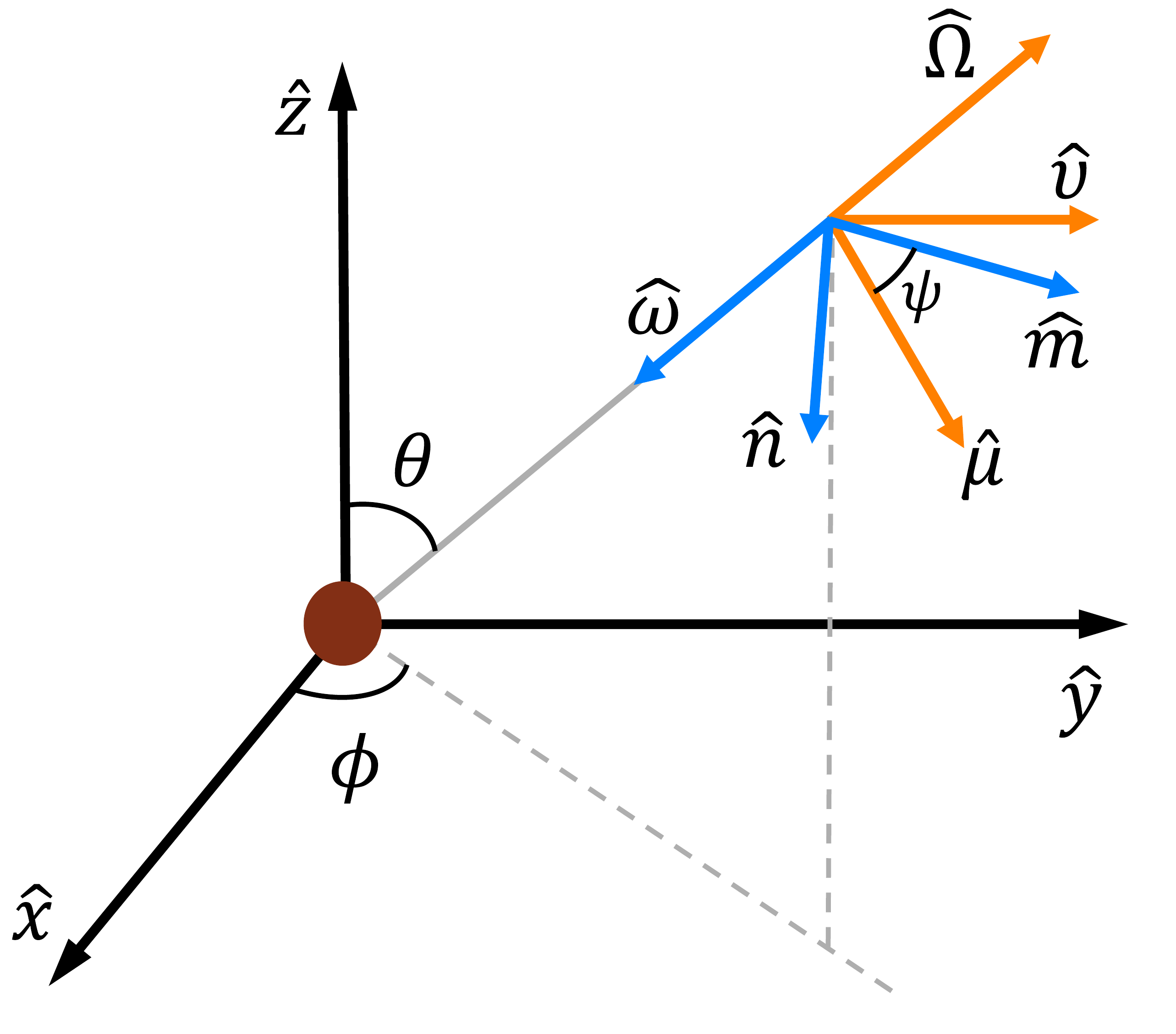}
    \caption{Source frame in the heliocentric coordinate system.}
    \label{fig:source_frame}
\end{figure}

In the heliocentric coordinate system, we naturally have an orthogonal basis $(\hat x, \hat y, \hat z)$. 
For a wave source in the direction $\hat \Omega(\theta, \phi)$, one can define a new set of orthogonal bases $(\hat u, \hat v, \hat \Omega)$, following the approach used in ground-based detectors~\cite{poissonGravityNewtonianPostNewtonian2014, Nishizawa:2009bf}: 
\begin{align}
    \begin{aligned}
        \hat u &= (\cos \theta \cos \phi, \cos \theta \sin \phi, -\sin \theta), \\
        \hat v &= (-\sin \phi, \cos \phi, 0), \\
        \hat \Omega &= \hat u \times \hat v = (\sin \theta \cos \phi, \sin \theta \sin \phi, \cos \theta).
    \end{aligned}
    \label{eq:uvOmega}
\end{align}

To obtain the orthogonal basis $(\hat u, \hat v, \hat \Omega)$, one can first rotate the $(\hat x, \hat y, \hat z)$ coordinate system by an angle $\phi$ around the $z$-axis, and then rotate it by an angle $\theta$ around the new $y$-axis, as shown in Fig.~\ref{fig:source_frame}. 

In the source frame, one can define a set of orthogonal bases $(\hat m,\hat n,\hat \omega)$, vector $\hat \omega$ is the direction of gravitational wave propagation, $(\hat m,\hat n)$ represents the transverse basis of the source frame, as shown in  Fig.~\ref{fig:source_frame}. 
The relationship between this set and the previously defined $(\hat u,\hat v,\hat \Omega)$ is given by~\cite{Yunes:2013dva, Apostolatos:1994mx}:
\begin{align}
    \begin{aligned}
        \hat m &= \hat u \cos \psi + \hat v \sin \psi, \\
        \hat n &= \hat u \sin \psi - \hat v \cos \psi, \\
        \hat \omega &= \hat m \times \hat n = -\hat \Omega.
    \end{aligned}
    \label{eq:mnOmega}
\end{align}

Then, six polarization tensors could be expressed in the heliocentric coordinate system as:
\begin{align}
    \begin{aligned}
        {e}_{ij}^{+} &= \hat m_i \hat m_j - \hat n_i \hat n_j, \quad
        {e}_{ij}^{\times} = \hat m_i \hat n_j + \hat n_i \hat m_j, \\
        {e}_{ij}^{x} &= \hat m_i \hat \omega_j + \hat \omega_i \hat m_j, \quad
        {e}_{ij}^{y} = \hat n_i \hat \omega_j + \hat \omega_i \hat n_j, \\
        {e}_{ij}^{b} &= \hat m_i \hat m_j + \hat n_i \hat n_j, \quad
        {e}_{ij}^{l} = \hat \omega_i \hat \omega_j.
    \end{aligned}
    \label{eq:polarization_tensors}
\end{align}

\subsection{Detector Configuration}

\begin{figure}
    \centering
    \subfigure[The standard Taiji-LISA network configuration in the most of literatures.]{
        \includegraphics[width=0.9\linewidth]{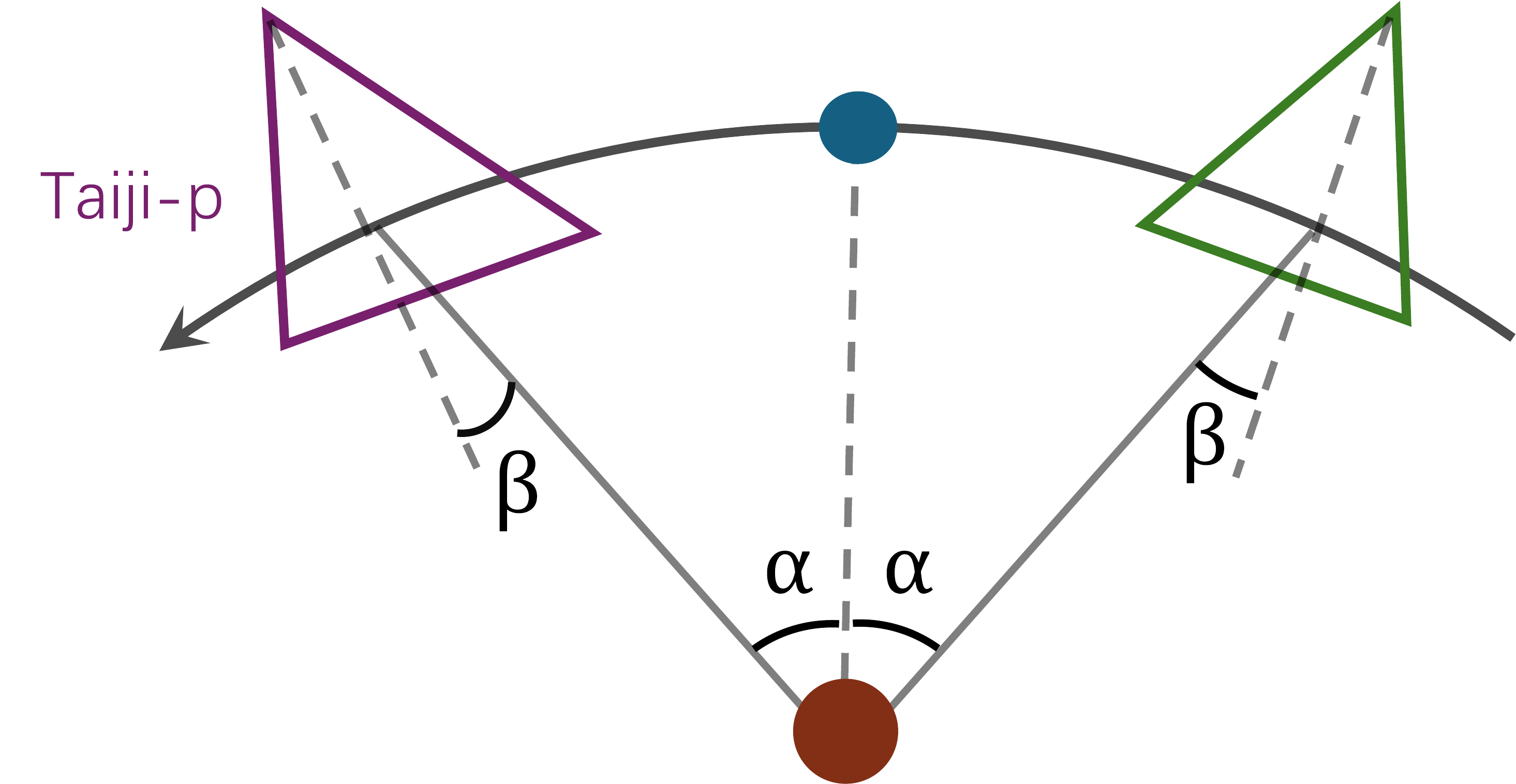}
        \label{fig:network_p}
        }
    \subfigure[The phase of three spacecrafts of Taiji are ahead of these of LISA by a fixed angle $\gamma$. 
    In other words, the three spacecrafts of Taiji rotate around the normal axis by an angle $\gamma$.]{
        \includegraphics[width=0.9\linewidth]{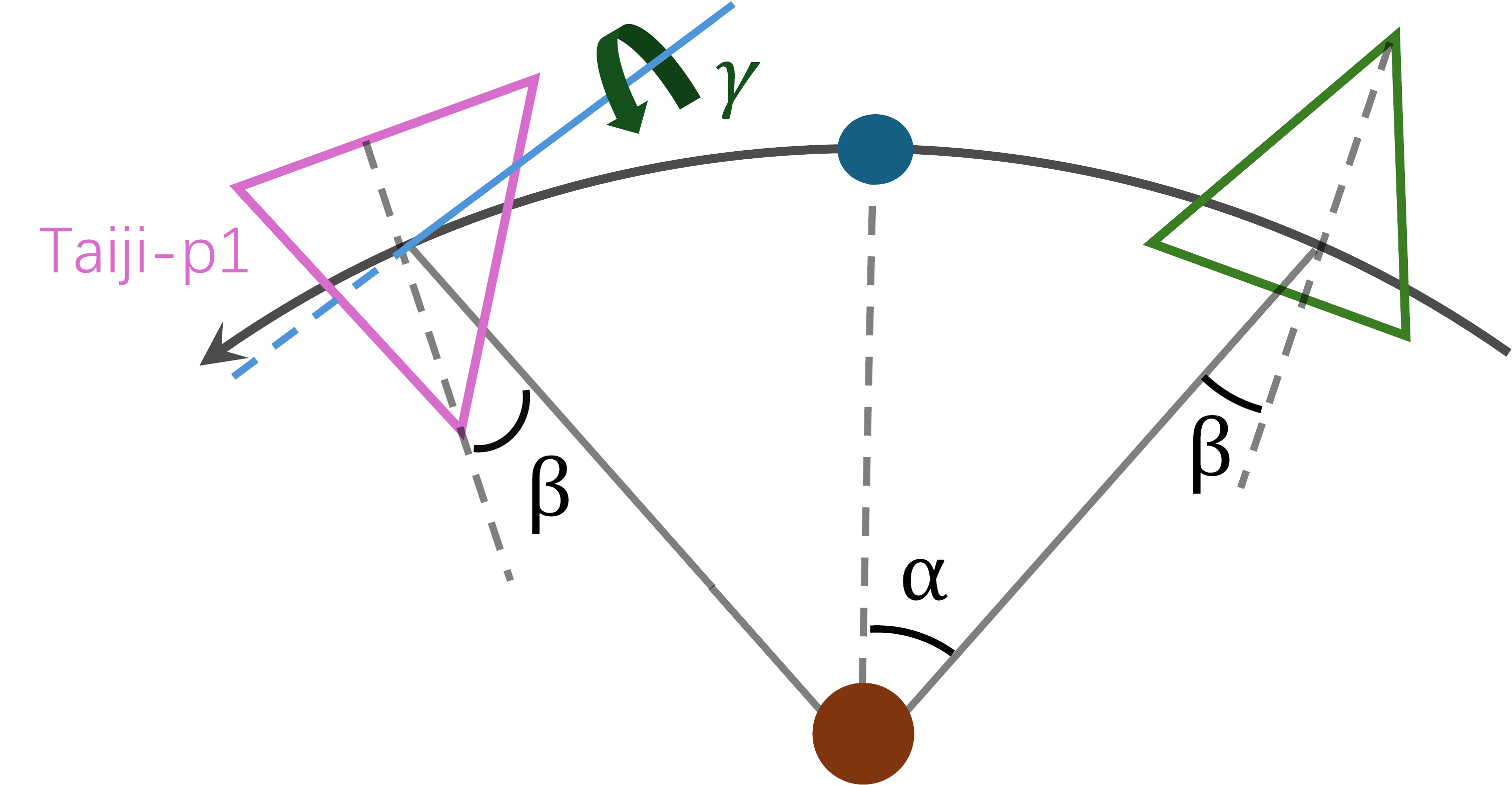}
        \label{fig:network_p1}
    }
    \caption{Two different network configurations for LISA and Taiji. 
    In this figure, $\alpha=20^\circ$ and $\beta=60^\circ$. 
    }
    \label{fig:network_p_p1}
\end{figure}

The geometric configurations of LISA and Taiji are shown in Fig.~\ref{fig:network_p_p1}. 
The three arms of LISA are each $2.5 \times 10^6$~km, while the arm length of Taiji is $3 \times 10^6$~km. 
The centers of mass of LISA and Taiji follow heliocentric trajectory, trailing $\pm 20^\circ$ ahead of Earth, respectively. The angle between the detector plane and the ecliptic plane is $\pm 60^\circ$, which is the requirement for orbital stability~\cite{dhurandharFundamentalsLISAStable2005}.

For Taiji and LISA, the detector plane can be oriented at $\pm 60^\circ$~\cite{shumanMassiveBlackHole2022}, this results in two possible configurations, which are named LISA-Taiji-p and LISA-Taiji-m~\cite{wangAlternativeLISATAIJINetworks2021}, respectively. When LISA and Taiji share the same ecliptic longitude, this specific alignment is designated as the LISA-Taiji-c configuration~\cite{wangAlternativeLISATAIJINetworks2021}. 
Different detector configurations perform differently for various scientific mission goals. 
Based on the results reported in the current literatures, the LISA-Taiji-c configuration has advantages in detecting the stochastic gravitational wave background~\cite{Wang:2022pav}, while LISA-Taiji-p and LISA-Taiji-m are more advantageous in detecting gravitational wave events~\cite{wangAlternativeLISATAIJINetworks2021}.

In this study, based on the LISA-Taiji-p configuration, we introduce a relative phase in the orientation between Taiji and LISA, terming this enhanced configuration as LISA-Taiji-p1($\gamma \neq 0$), Fig.~\ref{fig:network_p1}, which has not yet been applied in the study of testing non-tensor polarization modes. 
Conversely, the standard configuration of LISA-Taiji network (LISA-Taiji-p) has already been extensively studied~\cite{wuProspectsConstrainingPolarizations2024}, and this is also why we call it the standard configuration. 

{
Existing literature has noted that the limited alignment angles of the ground-based LIGO detectors resulted in suboptimal performance of the detector network~\cite{Hilborn:2018rio,Svidzinsky:2018hnx}. Therefore, including such a phase in space-based detectors is scientifically justified. 
} 

Following Refs.~\cite{Rubbo:2003ap, cutlerAngularResolutionLISA1998}, to the first order in orbital eccentricity, the coordinates $\mathbf{x}(t)$ of the three spacecrafts can be expressed in the heliocentric coordinate system as:
\begin{equation}
    \begin{aligned}
        x(t) &= R \cos \alpha + \frac{1}{2} e R [\cos(2\alpha - \beta) - 3 \cos \beta], \\
        y(t) &= R \sin \alpha + \frac{1}{2} e R [\sin(2\alpha - \beta) - 3 \sin \beta], \\
        z(t) &= -\sqrt{3} e R \cos(\alpha - \beta),
    \end{aligned}
    \label{eq:xyz}
\end{equation}
where $ R = 1 $ AU, $ e = L / (2 \sqrt{3} R) $ is the orbit eccentricity, $ \alpha = 2 \pi f_m t + \kappa $ with $ f_m = 1/\text{year} $, and $ \beta = 2 \pi n / 3 + \lambda $ (for $ n = 0, 1, 2 $). 
Here $ \kappa $ and $ \lambda $ are the initial ecliptic longitude and orientation of the spacecraft, respectively~\cite{Rubbo:2003ap, Liu:2020mab}; $\kappa=-20^{\circ}$ for Taiji and $\kappa=20^{\circ}$ for LISA; if $ \lambda=\gamma_0 $ for LISA, then one have $ \lambda=\gamma_0 $ for Taiji-p and $ \lambda=\gamma_0 + \gamma $ for Taiji-p1, as shown in  Fig.~\ref{fig:network_p_p1}. 
The direction from the spacecraft $ i $ to the spacecraft $ j $ is described by
\begin{equation}
    \hat{{r}}_{ij}(t) = \frac{\mathbf{x}_j(t) - \mathbf{x}_i(t)}{L}.
\end{equation}

\subsection{Signal Model}

Space-based detectors such as LISA and Taiji, when moving in heliocentric orbits, could introduce various modulation effects~\cite{Liu:2020nwz}. 
According to the results of Ref.~\cite{Rubbo:2003ap}, after applying the rigid adiabatic approximation, the interferometric results on spacecraft 1 can be expressed as: 
\begin{align}
    s_1(t) = \Re \Big(\mathbf{F} \big(t, f(\xi) \big) : \mathbf{H}(\xi) \Big),
\end{align}
where $\Re(X)$ means taking the real part of X; $\xi=t-\hat{k} \cdot \vec{x}$ represents the wavefront of the plane gravitational wave; $\mathbf{H}(\xi)$ is the tensor notation of $h_{ij}$ in Eq.~\eqref{eq:h_ij}; $\mathbf{a}:\mathbf{b}=a_{\mu\nu}b^{\mu\nu}$, and $\mathbf{F}\big(t, f(\xi)\big)$ is the antenna pattern function:
\begin{align}
    \begin{aligned}
        \mathbf{F}\big(t, f(\xi)\big) =& \frac{1}{2} \Bigg[ \Big(\hat{r}_{12}(t) \otimes \hat{r}_{12}(t)\Big) \mathcal{T}\Big(\hat{r}_{12}(t), f(\xi)\Big) \\
        & \hspace{0pt} - \Big(\hat{r}_{13}(t) \otimes \hat{r}_{13}(t)\Big) \mathcal{T}\Big(\hat{r}_{13}(t), f(\xi)\Big) \Bigg],
    \end{aligned}
\end{align}
where $\mathcal{T}$ is the transfer function~\cite{Rubbo:2003ap}, 
\begin{align}
    \begin{aligned}
        \mathcal{T}\big(\hat{r}_{ij}(t), f(\xi)\big) =& \frac{1}{2} \Bigg[ \text{sinc} \left( \frac{f(\xi)}{2f_*} 
        \Big(1 - \hat{r}_{ij}(t) \cdot \hat{k} \Big) \right)
        \\ & \hspace{-10pt} \times \exp \left( -i \frac{f(\xi)}{2f_*} \Big(3 + \hat{r}_{ij}(t) \cdot \hat{k}\Big) \right)
        \\ & \hspace{-25pt} + \text{sinc} \left( \frac{f(\xi)}{2f_*} \Big(1 + \hat{r}_{ij}(t) \cdot \hat{k}\Big) \right) 
        \\ & \hspace{-10pt} \times \exp \left( -i \frac{f(\xi)}{2f_*} \Big(1 + \hat{r}_{ij}(t) \cdot \hat{k}\Big) \right) \Bigg],
    \end{aligned}
\end{align}
where $\text{sinc}(X) \equiv \sin(X)/X$~\footnote{$\text{sinc}(X) \equiv \sin(\pi X)/(\pi X)$ 
in numpy, which is a package of Python.}, 
$f_* = 1/(2 \pi L)$ is the transfer frequency with $L$ being the arm length, one has $f_* = 0.019$ for LISA and $f_* = 0.0159$ for Taiji. 

We use data channels A and E for simplicity, which are linearly independent~\cite{Cornish:2003tz}.

\subsection{Noise}

The noise in a GW detector can be described by the one-sided 
noise power spectral density, denoted as $ S_n(f) $. 
For LISA, we use the noise curve in Refs.~\cite{LISACosmologyWorkingGroup:2019mwx, niuConstrainingScreenedModified2020}:
\begin{align}
    S_n(f) = \frac{4 S_{\text{acc}}(f) + S_{\text{other}}}{L^2} \left[ 1 + \left( \frac{f}{1.29 f_*} \right)^2 \right] + S_{\text{conf}}(f)
\end{align}
here $f_*$ and $L$ is the transfer frequency and arm length of LISA, respectively.
The acceleration noise $S_{\text{acc}}(f)$ is given by: 
    \begin{align}
        S_{\text{acc}}(f) =& \frac{9 \times 10^{-30} \, \text{m}^2 \text{Hz}^3}{(2\pi f)^4} \Bigg\{ 1 + \left( \frac{6 \times 10^{-4} \, \text{Hz}}{f} \right)^2
        \nonumber\\
        & \hspace{25pt} \times \left[ 1 + \left( \frac{2.22 \times 10^{-5} \, \text{Hz}}{f} \right)^8 \right] \Bigg\}.
    \end{align}
The other noise $S_{\text{other}}$ is given by:
\begin{align}
    S_{\text{other}} = 8.899 \times 10^{-23} \, \text{m}^2 \text{Hz}^{-1}.
\end{align}

In space-based gravitational wave detection, confusion noise mainly comes from the gravitational wave background produced by a large number of distant and unresolved astrophysical sources. 
These sources include, but are not limited to, binary white dwarf systems, binary neutron stars, and binary black holes.
Due to the huge number of these systems and the relatively weak gravitational wave signals they emit, individual signals are difficult to clearly identify and separate. 
As a result, they collectively contribute to a background that resembles noise. 
The confusion noise from unresolved binaries is given by~\cite{niuConstrainingScreenedModified2020, Robson:2018ifk, Cornish:2017vip}:
\begin{equation}
    S_{\text{conf}}(f) = \frac{A}{2} e^{-s_1 f^\alpha} f^{-7/3} \left\{ 1 - \tanh \Big[ s_2 \times \big(f - \kappa \big) \Big] \right\},
    \label{eq:noise_conf}
\end{equation}
with $ A = \left(\frac{3}{20}\right) 3.2665 \times 10^{-44} \, \text{Hz}^{4/3} $, $ s_1 = 3014.3 \, \text{Hz}^{-\alpha} $, $ \alpha = 1.183 $, $ s_2 = 2957.7 \, \text{Hz}^{-1} $, and $ \kappa = 2.0928 \times 10^{-3} \, \text{Hz} $.

For Taiji, we use the noise curve in Refs.~\cite{Cornish:2001qi, liuExploringNonsingularBlack2020}:
\begin{equation}
    S_n(f) = \frac{S_x}{L^2} + \frac{4 S_a}{(2 \pi f)^4 L^2} \left(1 + \frac{10^{-4} \text{Hz}}{f}\right) + S_{\text{conf}}(f)
\end{equation}
where $S_x = 64 \times 10^{-24}\mathrm{m}^2/\text{Hz}$, and $S_a = 9 \times 10^{-30}\mathrm{m^2s^{-4}}/\text{Hz}$. For $S_{\text{conf}}(f)$ in Taiji, we keep the same expression as LISA in Eq.~\eqref{eq:noise_conf}.


\section{Results}
\label{sec:results}

\begin{figure*}[htbp]
    \subfigure[the blue solid line represents the results from this work, while the red dashed line shows the results obtained when only two parameters are estimated.]{
        \includegraphics[width=0.45\linewidth]{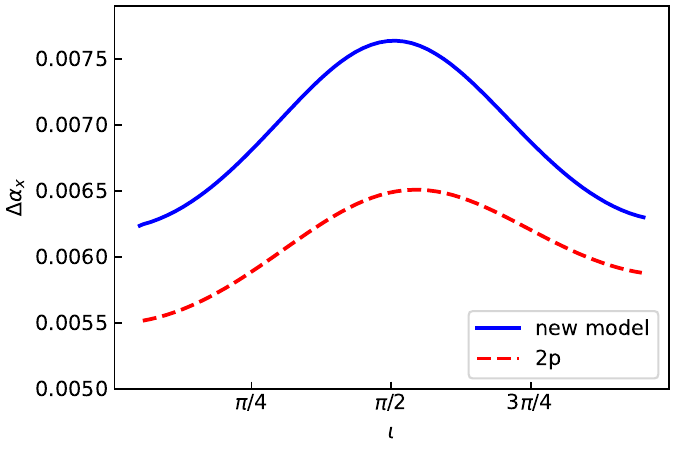}
        \label{fig:alpha_v_iota}
    }
    \subfigure[The result from reference~\cite{Liu:2020mab}, which has an observation time of 60 days. We have magnified and displayed the left and right ends of the figure in the subplot above the main figure (the orange solid line).]{
        \includegraphics[width=0.45\linewidth]{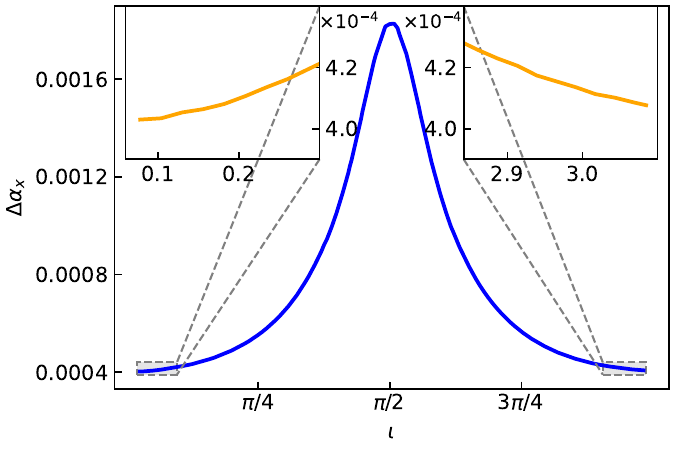}
        \label{fig:alpha_v_iota0}
    }
    \subfigure[$\Delta \alpha_x$ varies with polarization angle $\psi$.]{
        \includegraphics[width=0.45\linewidth]{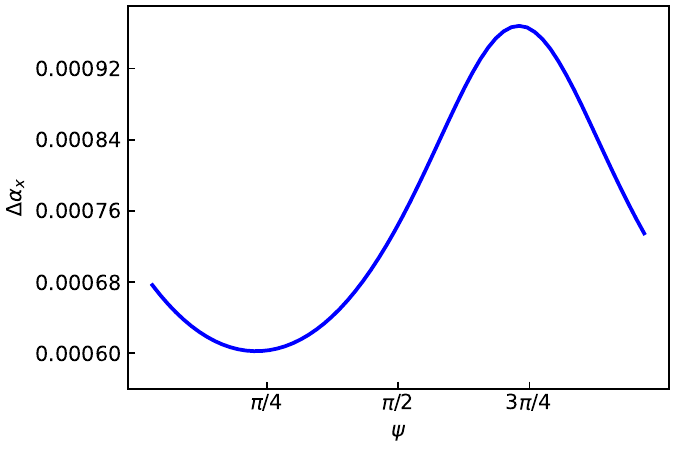}
        \label{fig:alpha_v_psi}
    }
    \subfigure[$\Delta \alpha_b$ varies with inclination angle $\iota$.]{
        \includegraphics[width=0.45\linewidth]{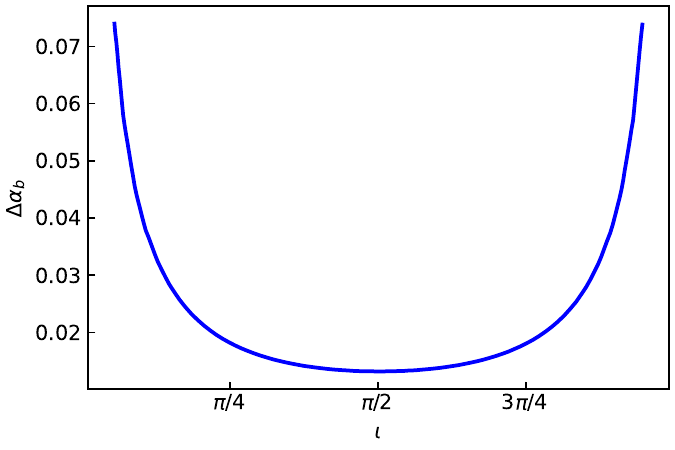}
        \label{fig:alpha_b_iota}
    }
    \caption{The 1$\sigma$ width of $\alpha_x$ and $\alpha_b$ 
    ($\Delta \alpha_x$ and $\Delta \alpha_b$) varies with inclination angle $\iota$ and polarization angle $\psi$. 
    In this result, we fixed $\alpha_x = \alpha_b = \alpha_l = 0$ and used the Taiji detector.
    }
    \label{fig:alpha_v}
\end{figure*}

There are already many software packages available for GW data analysis, such as PyCBC and Bilby~\cite{Biwer:2018osg,ashtonBilbyUserfriendlyBayesian2019}. 
The code used in this work is based on that shared by Ref.~\cite{Hu:2020rub} on \href{https://github.com/Li-mz/bilby/tree/SpaceInterferometer}{GitHub repository}, which is modified based on Bilby. 
The sampler PyMultiNest~\cite{buchnerXraySpectralModelling2014} is used to enable multi-core sampling acceleration. 
Since Bilby has a built-in interface for the Fisher information matrix, we can use the same code to perform both Bayesian estimation and Fisher estimation.

To study the detection capabilities of Taiji and LISA for non-tensor modes, we focus on the parameter estimation of three parameters, $\{ \alpha_x, \alpha_b, \alpha_l \}$. 
Their fiducial values may be chosen differently in various scenarios, and these will be explicitly specified in the specific results. 
Other parameters can be fixed in advance, $ 
M_1 = 1 \times 10^{5} \, M_{\odot}, 
M_2 = 1 \times 10^{5} \, M_{\odot}, 
\iota = \pi/4, \psi = 0.1, 
\theta = \pi/4, \phi = \pi, 
\alpha_d = 0.001, \alpha_q=96/5
$, and we choose luminosity distance $D_L=6790~\mathrm{Mpc}$ corresponding to a redshift $z \approx 1$.

We start with a time-domain signal, with a cutoff time corresponding to the ISCO time, and a total duration of 30 days. 
The advantage of doing this is that the frequencies of the non-tensor modes naturally turn out to be half of those of the tensor modes, including the cutoff frequency, in other words, the frequency sequence of the tensor modes is $ f_T \in [f_{min}, f_{\text{ISCO}}] $ and the frequency sequence of the non-tensor modes is $ f_{NT} \in \frac{1}{2}[f_{min}, f_{\text{ISCO}}] $. 
After applying the Fourier transform, one can obtain the waveforms in the frequency domain.
However, in the previous works, the frequency domain waveforms are obtained using the stationary phase approximation~\cite{Liu:2020mab, wuProspectsConstrainingPolarizations2024}.

\subsection{Inclination angle}
\label{sec:sub:iota}

In this subsection, we calculate how the parameter uncertainties vary with $\iota$. As the observation time decreases (compared with Ref.~\cite{Liu:2020mab}), the parameter uncertainties become broader, 
as shown in Fig.~\ref{fig:alpha_v_iota}. 
Moreover, we note that $\Delta \alpha_x$ no longer appears to be as symmetric, as shown in Fig.~\ref{fig:alpha_v_iota}. 
As a verification, we have placed the results from Ref.~\cite{Liu:2020mab} in Fig.~\ref{fig:alpha_v_iota0} (the blue solid line), and magnified the curves at the left and right ends (the orange solid line in the subplots at 
the top of figure); indeed, we observe the asymmetry on both sides. 
When $\iota=\pi$, the value of $\Delta \alpha_x$ is approximately 2\% larger than that when $\iota=0$. 
If we estimate only the two parameters $\alpha_x$ and $\alpha_b$, this asymmetry becomes even more pronounced, see the red dashed line in Fig.~\ref{fig:alpha_v_iota}.

\begin{figure*}[htbp]
    \includegraphics[width=\linewidth]{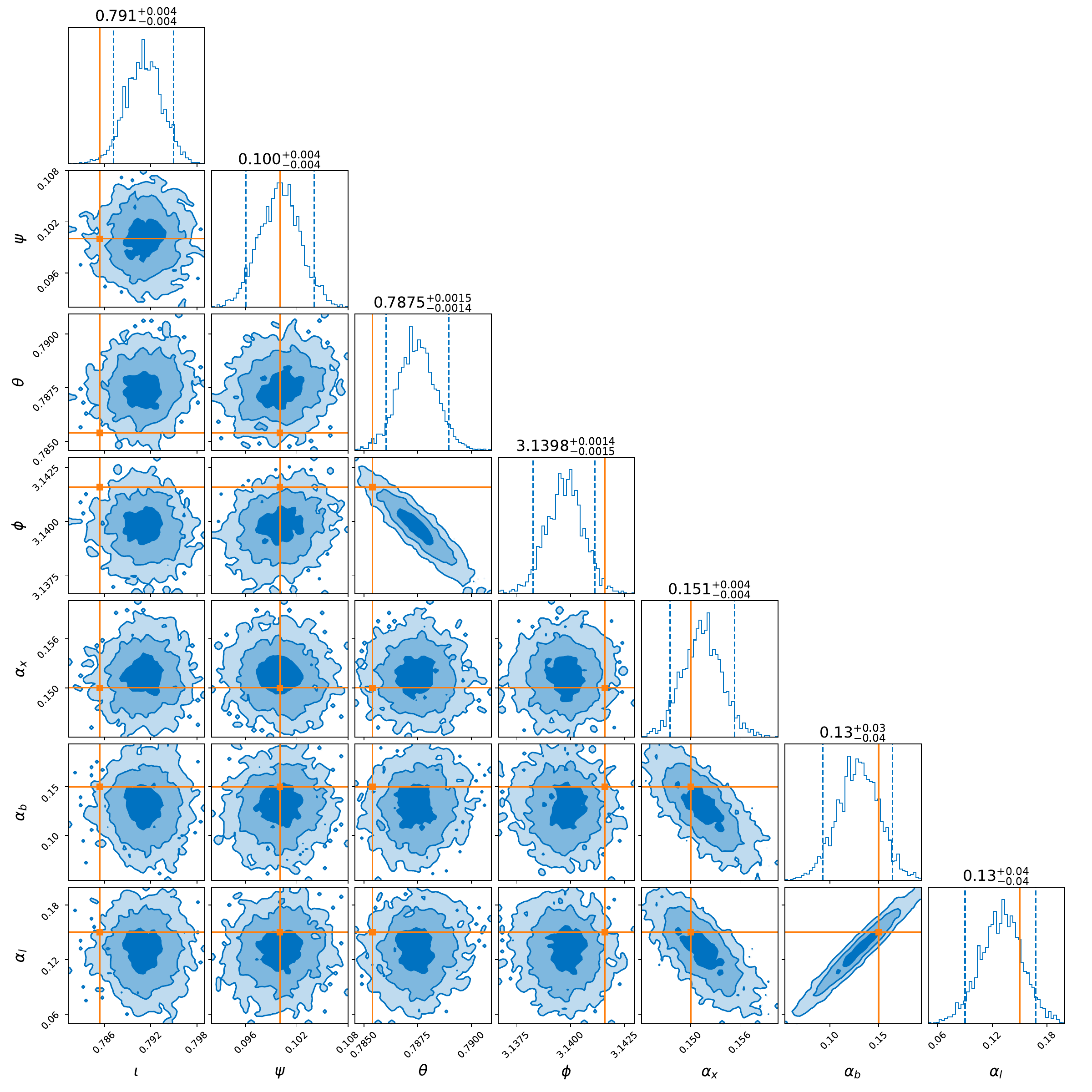}
    \caption{Corner plot illustrating the posterior distributions and correlations among various parameters in a multi-dimensional analysis. 
    The contour plots depict the joint probability distributions between pairs of parameters, while the line plots along the diagonal show the marginal distributions for each parameter.
    The orange lines indicate the injected value, and the blue shaded regions represent the 95\% confidence intervals. 
    This result based on the Taiji detector. 
    }
    \label{fig:corner}
\end{figure*}

This asymmetry has not been mentioned in any of the previous studies~\cite{Liu:2020mab, wuProspectsConstrainingPolarizations2024}, probably due to three perspectives. 
First, it is due to the difference in signals between $ \iota = 0 $ and $ \iota = \pi $. 
In Eq.~\eqref{eq:waveforms_ppE_2}, $ h_x $ is proportional to $ \cos \iota $, so one has 
$$ h_x \big | _{\iota=0} = - h_x \big | _{\iota=\pi}, $$
but, 
$$ h_y \big | _{\iota=0} = h_y \big | _{\iota=\pi}, $$
it is precisely the inconsistency in their behavior that leads to the difference in signals between $ \iota = 0 $ and $ \iota = \pi $. 
This is the fundamental source of the asymmetry. 
Second, even if such asymmetry exists, it is not easily noticeable because the longer the observation duration, the larger the ratio of the width at $ \iota = \pi/2 $ to that at $ \iota = 0 $. In Fig.~\ref{fig:alpha_v_iota0}, this ratio is more than a factor of four, making the 2\% difference relatively insignificant and difficult to observe. 
Third, there is a strong correlation between parameters $\alpha_x, \alpha_b, \alpha_l$, as shown in Fig.~\ref{fig:corner}. 
The widths of $\alpha_b$ and $\alpha_l$,
$\Delta \alpha_b$ and $\Delta \alpha_l$, are much larger, approximately 10 times that of $\alpha_x$, $\Delta \alpha_x$, as shown in Fig.~\ref{fig:alpha_b_iota}. 
When scalar modes are present, they can affect the performance of the vector mode ($\Delta \alpha_x$).

However, the variation of $\Delta \alpha_x$ with the polarization angle $\psi$ is consistent with the behavior reported in Ref.~\cite{Liu:2020mab}. 
And $\Delta \alpha_x$ is equal at $\psi = 0$ and $\psi = \pi$, as shown in Fig.~\ref{fig:alpha_v_psi}, because of the behavior of Eq.~\eqref{eq:polarization_tensors}: 

\begin{align}
    e^x_{ij} \big |_{\psi=0} = -e^x_{ij} \big |_{\psi=\pi}, 
    \nonumber\\
    e^y_{ij} \big |_{\psi=0} = -e^y_{ij} \big |_{\psi=\pi}.
    \nonumber
\end{align}

The overall sign reversal of the polarization tensor does not lead to differences in the parameter estimation of gravitational wave signals. 
The behavior of the other parameters is consistent with that reported in Ref.~\cite{Liu:2020mab}; therefore, we will omit the detailed discussion of this part here and refer the reader to Ref.~\cite{Liu:2020mab} for more details.

\subsection{Bayesian posterior distribution}

In Fig.~\ref{fig:corner}, we present the results of Bayesian parameter estimation, as existing literature commonly uses the Fisher information matrix to reduce computational costs. 
In this figure, one can see that $\alpha_x$ is negatively correlated with both $\alpha_b$ and $\alpha_l$. 
On the other hand, $\alpha_b$ and $\alpha_l$ are positively correlated, and the correlation is relatively strong. 
On ground-based detectors, it is not possible to distinguish between $\alpha_b$ and $\alpha_l$ ($ F_b = -F_l $ in Ref.~\cite{poissonGravityNewtonianPostNewtonian2014}), 
which has led to hopes for space-based detectors. 
However, in this study, we assume that the non-tensor modes are sourced by quadrupole radiation, and combined with our truncation at the ISCO (innermost stable circular orbit), the frequency range of the non-tensor modes still falls within the low-frequency regime, making it difficult to effectively distinguish between $\alpha_b$ and $\alpha_l$. 
If the waveform of the merger and ringdown phases of the non-tensor modes can be obtained, it may help distinguish between $\alpha_b$ and $\alpha_l$.

\subsection{Network}

\newcommand{\mywidth}{0.3}

\begin{figure}[tb]
\centering
    \begin{overpic}[width=0.9\linewidth]{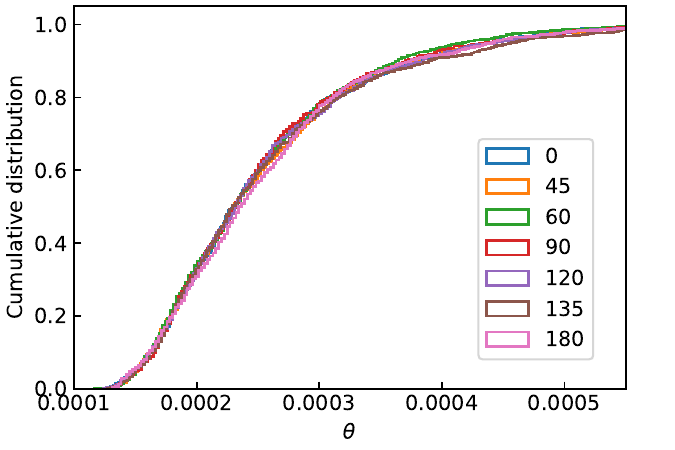}
        \put(14, 60){\bf{(a)}}
    \end{overpic}
    \begin{overpic}[width=0.9\linewidth]{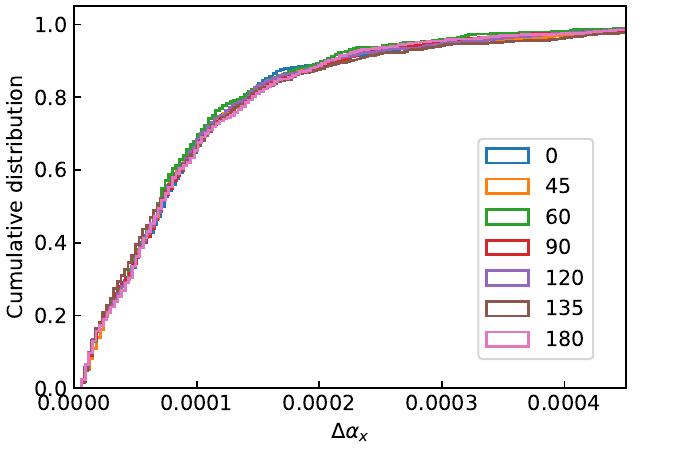}
        \put(14, 60){\bf{(b)}}
    \end{overpic}
\caption{Cumulative distribution of the average uncertainties of 1800 sources in parameters $\theta$ and $\alpha_x$ with different $\gamma$ values. 
The horizontal axis denotes the uncertainty associated with the parameters, while the vertical axis corresponds to the cumulative distribution probability. It means the configuration 
of LISA-Taiji-p if $\gamma=0$. 
The results show that the uncertainty distributions under all $\gamma$ values are not significantly different. The behaviors of the remaining parameters are consistent with the results presented in the figure. Hence, networks with different $\gamma$ values exhibit very similar performance.}
\label{fig:gamma}
\end{figure}

In this work, we also study the capability of a new configuration of LISA and Taiji network, as shown in Fig.~\ref{fig:network_p1}, in detecting non-tensor modes. 
This configuration is similar to the hexagram configuration used in Japan's DECIGO~\cite{Kawamura:2008zza}, except that there is a $40^\circ$ angle between Taiji and LISA. 
This configuration differs from the original one by only one additional parameter, a relative phase $\gamma$. 
However, to the best of our knowledge, there are no existing discussions on this specific configuration in the literature related to detector networks. 
Therefore, our work fills this gap in testing 
non-tensor modes. 

The orientation of the spacecraft $\lambda$ in Eq.~\eqref{eq:xyz} is set to 0 for LISA and $\gamma$ for Taiji. 
In principle, we could treat $\gamma$ as a free parameter to calculate the sensitivity curves. 
However, in current treatments of non-tensor polarization modes, either no consideration or only first-generation Time-Delay Interferometry are taken into account, the reliability of the sensitivity curves remains to be verified. 
Moreover, the choice of which configuration of LISA-Taiji network will be used is still under discussion. In this context, our results are sufficiently meaningful in demonstrating the 
feasibility of such an approach.

In this paper, we calculate the parameter inference precision with different values of $\gamma$, averaged over 1800 sources uniformly distributed in the sky($\theta$, $\phi$), and random  distributed in ($\iota$, $\psi$). 
In Fig.~\ref{fig:gamma}, the cumulative distributions of the parameters $\Delta \theta$ and $\Delta \alpha_x$ are shown. The curves for different $\gamma$ values overlap closely, indicating that, under all-sky averaging, the initial phase differences associated with different $\gamma$ values have no significant impact on the detector's performance. The behaviors of the other parameters ($\Delta \phi$, $\Delta \alpha_b$, $\Delta \alpha_l$) are consistent with those displayed in Fig.~\ref{fig:gamma}, and thus are not shown individually.

This observation is also reasonable, as the independent A and E data channels in LISA-like space detectors are mathematically equivalent to two L-shaped interferometers rotated by 45 degrees relative to each other~\cite{setoMeasuringParityAsymmetry2020}.  
Consequently, a rotation of the space detector about the normal axis of its detector plane corresponds to a linear transformation of the A and E channels~\cite{setoMeasuringParityAsymmetry2020}. 
Hence, the essential factor in networking LISA-like detectors is the angular separation between their respective detector planes, rather than the relative orientation within a single fixed plane.

\section{Summary and Outlook}
\label{sec:summary}

This paper systematically investigates the potential of space-based gravitational wave detectors, such as the LISA-Taiji network and the Taiji single detector, in detecting non-tensor polarization modes. 
We use a new detector network configuration to calculate the accuracy of parameter estimation for non-tensor modes. 
Using Bayesian inference and Fisher information matrix, we analyze the vector and scalar polarization components in gravitational wave signals from the inspiral of supermassive binary black holes, and compare the sensitivity performance across different detector configurations. 

Our main findings include: 
First, 
we start from the time-domain signal and apply the Fourier transform to avoid 
using the stationary phase approximation. When reducing the observation time, an increase in the width of the parameter uncertainty is found.

Second, 
asymmetry in the estimation of the non-tensor mode parameter $\alpha_x$: 
In the previous works, the non-tensor mode parameter $\alpha_x$ displays symmetric behavior with respect to inclination angle $\iota = 0$ and $\iota = \pi$. 
However, through our signal modeling, we find that this symmetry does not strictly hold, especially when the observation duration is relatively short. 
This asymmetry arises from the different dependence of the waveform on the orbital inclination angle $\iota$. 
However, the observation duration and the correlation between the parameters affects whether this phenomenon is easily observable. 

Third, the strong correlations between parameters limit the ability to estimate them independently.
Bayesian posterior distributions reveal significant covariance among the non-tensor parameters, particularly a strong positive correlation between $\alpha_b$ and $\alpha_l$.
This indicates that, under the current low-frequency signal model, it is difficult to effectively distinguish between these two scalar modes.
At the same time, due to the assumption that non-tensor modes are dominated by dipole radiation, the frequency distribution of non-tensor modes tends more toward the low-frequency range compared to tensor modes, which also increase the difficulty of distinguishing between them, 
since the transfer function $\mathcal{T}$ can be approximated as 1 at low frequencies. 
In the future, a ppE framework that includes the merger and ringdown phases may help resolve this issue.

Fourth, detection capability of the new detector network configuration: 
We use a new LISA-Taiji network configuration ($\gamma \neq 0$), in which the Taiji detector has a fixed phase offset relative to LISA. 
Numerical results show that the new configuration has the same detection performance as LISA-Taiji-p. This indicates that the angular separation between detector planes is the key factor determining the performance of a space detector network, rather than the relative orientation within a single detector plane as considered in this work.

To further advance the study of non-tensor polarization modes, the following directions are worth deeper exploration.
First, more complete waveform models:
Current studies mainly rely on approximate waveforms based on the inspiral phase. 
Incorporating merger and ringdown waveforms in future analyses would help better distinguish between different polarization modes and significantly improve the accuracy of parameter estimation. 
Second, although the performance of different configurations is comparable in this study, the high-frequency region and the case of unequal arm lengths have not yet been considered and require further investigation. 
Third, impact of data gaps:
Data gaps can lead to a reduction in signal-to-noise ratio, affecting the precision of parameter estimation. 
The impact of data gaps on the detection of non-tensorial polarization modes is not yet fully understood and needs further investigation.

\begin{acknowledgments}
The work of Jun-Shuai Wang was supported in part by the National Key Research and Development Program of China (No.~2020YFC2201300), the National Natural Science Foundation of China (NSFC) under Grant No.~12347103 and the Fundamental Research Funds for the Central Universities.
Chang Liu is supported by the National Natural Science Foundation of China under Grant No. 12405074.
\end{acknowledgments}

\bibliographystyle{main-apsrev4-2}
\bibliography{main}
\end{document}